\documentclass[twocolumn,aps,prb]{revtex4}   
\usepackage{amsmath,amssymb}    
\usepackage{graphicx}   
\draft

\begin{document}

\title{Falling chains}
\author{Chun Wa Wong}
\email{cwong@physics.ucla.edu}
\author{Kosuke Yasui}
\affiliation{Department of Physics and Astronomy, 
University of California, Los Angeles, California 90095-1547}


\begin{abstract}
The one-dimensional fall of a folded chain with one end suspended 
from a rigid support and a chain falling from a resting heap on a table 
is studied. Because their Lagrangians contain no explicit time dependence,
the falling chains are conservative systems. Their equations of motion are
shown to contain a term that enforces energy conservation when masses are
transferred between subchains. We show that Cayley's 1857 energy
nonconserving solution for a chain falling from a resting heap is incorrect
because it neglects the energy gained when a transferred link leaves a 
subchain. The maximum chain tension measured by Calkin and March for the 
falling folded chain is given a simple if rough interpretation. 
Other aspects of this falling folded chain are briefly discussed.
\end{abstract}

\maketitle

\section{Introduction}

A folded flexible heavy chain is suspended from a rigid support
by its two ends placed close together. One end is then released 
in the  manner of a bungee fall, while the stationary arm gets 
longer. Calkin and March have noted that the system is conservative, 
``there being no dissipative mechanisms.''\cite{Calkin89} Energy 
conservation allows them to describe the one-dimensional motion of 
the falling chain simply and transparently in the continuum 
limit where the link length goes to zero: As the chain falls, energy 
conservation concentrates the entire mechanical energy in the still 
falling arm. When the mass of the falling arm finally vanishes at the 
end of the fall, both its falling velocity $v$ and its falling 
acceleration diverge to infinity.\cite{Calkin89} This phenomenon of 
energy concentration is similar to that occurring in the crack of 
the whip. In the words of Bragg\cite{Bragg20,Goriely02}, 
a shock ``wave runs down the cord and carries energy to the lash at 
the end,'' where the velocity diverges to 
infinity in the continuum limit\cite{Kurcharsky41,Rosenberg77}.

Calkin and March\cite{Calkin89} went on to measure the 
falling motion of an actual chain 2 m long containing 81 links. They 
found that the physical chain does indeed fall faster than free fall,
and that the continuum model accurately describes the experimental chain 
motion except near the end of the chain fall. Their meansurement of the 
chain tension $T$ at the fixed support of the chain is particularly
interesting. The theoretical chain tension given by the continuum model
contains a term proportional to $v^2$ of the falling velocity. It therefore 
increases without limit as the theoretical value of $v$ becomes infinite 
at the end of the fall. Calkin and March\cite{Calkin89} found that the 
experimental tension increases 
only to a maximum value of about 25$Mg$, where $M$ is the total mass of 
the chain. This maximum tension is of course far in excess of the maximum 
value of only 2$Mg$ expected when the falling end is falling freely, thus
demonstrating beyond doubt that the folded chain indeed falls faster than 
$g$. We shall explain in Sect.\ref{sect:tension} that it is the finite 
size of the link that prevents $T$ from going to the infinite 
value predicted by the continuum model.

In many older textbooks on mechanics,\cite{Love21,Lamb23,Chester80} the
falling arm is incorrectly described as freely falling and is brought to
rest by inelastic impacts at the fold of the chain. The kinetic energy
loss in a completely inelastic collision is real. It was first described 
by Lazare Carnot,\cite{Sommerfeld52,Ball60,Gillispie71} father of the Sadi
Carnot of thermodynamics. The effect is called Carnot's energy loss or
Carnot's theorem in Sommerfeld's book on mechanics.\cite{Sommerfeld52} 
The effect of an impulse alone on a dynamical system was treated correctly
by an additional term by Lagrange.\cite{Lagrange97,Whittaker44} In contrast,
Hamel\cite{Hamel49} obtained the correct solution for the falling chain by 
assuming energy conservation. We shall show that energy conservation results 
because the Carnot energy loss caused by a transferred mass absorbed by the 
receiving subchain is counterbalanced by the energy gained when the mass 
leaves the ``emitting'' subchain.

The Calkin-March observation\cite{Calkin89} that the folded chain 
falls faster than $g$ was subsequently confirmed
experimentally by Schagerl {\it et al.}\cite{Schagerl97} Photographic
evidence can also be found in Ref.~\onlinecite{Taft}. The correct solution
of the motion of the falling folded chain by energy conservation has been
included in some recent textbooks on classical
dynamics.\cite{Marion95,Meriam02,Thornton04}. 

Schagerl {\it et al.}\cite{Schagerl97} were unaware of the measurement of
Calkin and March.\cite{Calkin89} The results of their 
measurements\cite{Schagerl97} came as a surprise to them because they had 
concluded by
theoretical arguments that the chain fell only as fast as $g$, and that
the total mechanical energy was not conserved.\cite{Steiner95,Crellin97} In
these earlier papers, the authors rejected Hamel's energy conserving
solution,\cite{Hamel49} and claimed that there was dissipation caused by
the inelastic but momentum-conserving impacts at the fold of the chain.
They justified their treatment by citing Sommerfeld's use of Carnot's
energy loss in another falling chain problem,\cite{Sommerfeld52} which we
will describe in the following. 

The experimental observation\cite{Schagerl97} that the free end of the 
falling folded chain falls faster than $g$ might have led the authors
of Ref.~\onlinecite{Schagerl97} to conclude that the motion of the falling
chain is non-unique, because ``it is important to note that for the folded
string itself there exist more solutions which fulfill the balance of
linear momentum (but do not conserve the mechanical
energy).''\cite{Schagerl97} This non-uniqueness is the paradox referred to
in the title of their paper.\cite{Schagerl97}

The conclusion that non-unique solutions exist is clearly untenable because 
whether the chain is energy-conserving or not, its equation of motion is a 
linear differential equation with a unique solution for a given set of
initial conditions. Hence the experimental 
observation\cite{Calkin89,Schagerl97,Taft} of faster than $g$ fall proves 
that the motion cannot be the freely falling, energy nonconserving one. 
Thus there is no paradox. 

A review article by Irschik and Holl\cite{Irschik04} mentions the same
erroneous interpretation that for the falling folded chain, momentum is
conserved but mechanical energy is not conserved. These authors knew of the
experimental work in Ref.~\onlinecite{Schagerl97} but not that of
Ref.~\onlinecite{Calkin89}. In a previous paper on Lagrange's equations,
Irschik and Holl\cite{Irschik02} were puzzled by the result of 
Ref.~\onlinecite{Schagerl97}
because they thought that the string tension at the base of the falling arm
($N$ in their Eq.~(6.22)) should vanish, and therefore the arm
should fall freely. They realized that this conclusion is not consistent with
the observation of Ref.~\onlinecite{Schagerl97}. 

We shall show that the erroneous conclusion of energy loss comes from the 
neglect of the energy gained when the transferred mass at the fold of the 
chain leaves the falling arm. This energy gain is the time reverse of the 
Carnot energy loss incurred when the transferred mass is received by the 
stationary arm of the folded chain. 

There is another falling chain problem for which the consensus 
is that the 
total mechanical energy is not conserved. The steady fall of a stationary
chain resting on a table link by link through a hole in the table
appears to have been first studied by Arthur Cayley in 
1857.\cite{Cayley57,Irschik04} He treated the motion as a continuous-impact 
problem leading to a nonconservative system and a acceleration 
of $g/3$. Cayley's falling chain problem appears as Problem I.7 of 
Sommerfeld,\cite{Sommerfeld52} where the connection to Carnot's energy 
loss is explicitly stated. It can also be found in 
Refs.~\onlinecite{Love21a,Lamb23a,Saletan71,Chicon03,Marion95p,Thornton04p,Meriam02}. 
Note that Problem 9-15 in Ref.~\onlinecite{Marion95p} has been rewritten in 
Ref.~\onlinecite{Thornton04p} without any mention of energy loss. However,
the solutions given in the instructor's
manuals\cite{Marion95IM,Thornton04IM} are identical.

The only dissent we have found of this common consensus that energy is 
not conserved is in the recent paper by de Sousa and Rodrigues.\cite{Sousa04}
They first describe the falling folded chain by using a Newton equation for 
the two variable mass subchains that contains the gravitational force but 
no chain tension. They obtain the wrong or 
energy-nonconserving solution with acceleration $a=g$. They then solve the 
problem of the chain falling from a resting heap in a different way by 
assuming energy conservation. This assumption yields the right solution, as we 
shall show in the following. Their solution is the only correct solution 
we have been able to find in the literature for the chain falling 
from a heap on a table.  

In Sec.~\ref{sect:massX} we shall show specifically that the 
transfer of a link from the heap to the falling subchain is the same 
energy conserving process that operates in the falling 
folded chain, namely an exoergic mass emission followed by a 
counterbalancing endoergic mass absorption. We will see that 
Cayley and others considered only half of a two-step mechanical process that 
is energy-conserving as a whole.

Given the brief history of falling chains sketched here, it is 
interesting to determine unambiguously when a mechanical system such as a 
falling chain is energy conserving. The answer was already given in 1788 
by Lagrange.\cite{Lagrange97} In modern terminology 
using the Lagrangian $\mathcal{L}(x,v)$ and the Hamiltonian 
$\mathcal{H}(x,p)$, two conditions must be satisfied for the mechanical 
energy $E$ to be conserved: $E = \mathcal{H}$ and 
$\partial \mathcal{L}/\partial t = 0$. Consequently 
\begin{equation}
\frac{dE}{dt} = \frac{d\mathcal{H}}{dt} 
= -\frac{\partial \mathcal{L}}{\partial t} = 0,
\label{ECons}
\end{equation}
as we shall discuss in in Sec.~\ref{sect:E}. We shall 
also write the condition $E = \mathcal{H}$ in the original form given 
by Lagrange,\cite{Lagrange97} who referred to kinetic energies 
as ``live forces'' ({\it forces vives}). These
conditions are well known and can be found in most 
textbooks on analytical mechanics, but they have been too 
infrequently applied on actual physics problems. 

To show explicitly how this energy conservation enters in the mass 
transfer between subchains, we begin in Sec.~\ref{sect:massX} with 
the standard force equation of motion for a variable mass 
system\cite{Sommerfeld52,Thorpe62,Tiersten69,Sousa04,Matolyak90}. 
For the special case where no external forces act on these subchains, 
we show explicitly that the mass transfer is made up of an exoergic 
mass emission followed by an endoergic mass absorption when the 
transferred mass sticks inelastically to the receiving arm. We also
find that the complete process of mass transfer conserves 
mechanical energy when the transferred mass has the velocity given 
to it by Lagrange's equation of motion. Hence Lagrange's formulation
gives both the simplest and the most complete 
description of the motion of both falling chains.

There is an important practical difference between the two falling chains,
however. The link-by-link mass transfer of the falling folded chain 
is automatically guaranteed at the fold of the chain, but is difficult 
to realize for a real chain falling from a resting heap. The folded 
chain always falls in more or less the same way, but the motion of the 
resting heap depends on its geometry. More than one link at a time might be 
set into motion as the chain falls, and some of 
them might even be raised above the table before falling off it. These 
complications make it difficult to check the idealized theoretical result by 
an actual measurement. We therefore concentrate on the falling folded chain 
in the rest of the paper. In Sec.~\ref{sect:tension} we give a simple-minded 
interpretation 
of the maximal chain tension measured by Calkin and March.\cite{Calkin89}
Then we explain in Sec.~\ref{sect:last} how to understand the total loss 
of kinetic energy at the moment the chain reaches full extension, and why 
the chain rebounds against its support afterward. In Sec.~\ref{sect:conclusion} 
we pay tribute to Lagrange's formulation of classical mechanics.

\section{The Lagrangian and the Hamiltonian}
\label{sect:E}

\begin{figure}[h]
\begin{center}
\includegraphics{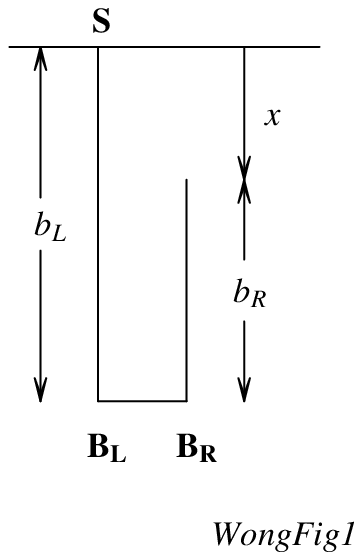}
\caption{\label{fig:bungee} The falling folded chain.}
\end{center}
\end{figure}

Figure~\ref{fig:bungee} shows the folded chain 
when its falling end has fallen a distance $x$. 
The chain is flexible, and has mass $M$, length $L$, and a 
uniform linear mass density $\mu = M/L$. Its Lagrangian in the idealized
one-dimensional treatment is
\begin{equation}
\mathcal{L}(x,v) = \frac{\mu}{4}(L-x)v^2 + MgX,
\label{L}
\end{equation}
where $v = \dot{x}$ and 
\begin{equation}
X = \frac{m_{L}x_{L} + m_{R}x_{R}}{M} 
= \frac{1}{4L}(L^2 + 2Lx - x^2)
\label{X}
\end{equation}
is its center of mass (CM) position measured in the downward direction. 
Here $m_L$ is the mass, $b_L$ is the length and $x_L$ is the CM position
of the left arm, while the corresponding quantities for the right arm are 
$m_R$, $b_R$ and $x_R$: 
\begin{subequations}
\label{mbx}
\begin{align}
m_i &= \mu b_i \\
b_{L} &= \frac{1}{2}(L+x)\\
b_{R} & = \frac{1}{2}(L-x) \\
x_{L} &= \frac{1}{4}(L+x) \\
x_{R} & = \frac{1}{4}(L+3x).
\end{align}
\end{subequations}
The parameters in the Lagrangian are time-independent and hence 
$\partial \mathcal{L}/\partial t = 0$.

The Hamiltonian of the falling folded chain is
\begin{equation}
\mathcal{H}(x,p_{R}) = p_{R} v - \mathcal{L}(x,v)
= \frac{p_{R}^2}{2m_{R}} - MgX = E.
\label{H}
\end{equation}
The canonical momentum,
\begin{equation}
p_{R} = \frac{\partial \mathcal{L}}{\partial v} = m_{R}v,
\label{pR}
\end{equation}
is the momentum of the right arm. Hence Eq.~(\ref{ECons}) is 
satisfied and the system is conservative.

The identity 
$d\mathcal{H}/dt = - \partial \mathcal{L}/\partial t$ used in 
Eq.~(\ref{ECons}) follows from the relation
\begin{equation}
\frac{\partial \mathcal{H}}{\partial x}\frac{dx}{dt} 
+ \frac{\partial \mathcal{H}}{\partial p}\frac{dp}{dt} = 0.
\label{PartH}
\end{equation}
These two terms cancel each other because the total time derivatives satisfy
the canonical equations of motion of Hamilton\cite{Hamilton35,Dugas55}
\begin{subequations}
\begin{align}
\frac{dx}{dt} &=\frac{\partial \mathcal{H}}{\partial p} \\
\frac{dp}{dt} &= - \frac{\partial \mathcal{H}}{\partial x}.
\label{CanEqs}
\end{align}
\end{subequations}

Equation~\eqref{ECons} can also be obtained
without using the Hamiltonian. We start with $E = 2K -\mathcal{L}$, where
$K$ is the kinetic energy, and write
\begin{equation}
\frac{dE}{dt} = \frac{d}{dt}(2K) 
- \Big( \frac{\partial \mathcal{L}}{\partial x} \dot{x} +
\frac{\partial \mathcal{L}}{\partial v} \dot{v} + 
\frac{\partial \mathcal{L}}{\partial t}\Big).
\label{dEdt}
\end{equation}
The second term on the right can be written in terms of 
$\partial \mathcal{L}/\partial v$ by using Lagrange's equation 
of motion\cite{Lagrange97a} 
\begin{equation}
\frac{\partial \mathcal{L}}{\partial x} = \frac{d}{dt}
\Big( \frac{\partial \mathcal{L}}{\partial v} \Big)
\label{Lagrange}
\end{equation}
to simplify $dE/dt$ to the form obtained by Lagrange\cite{Lagrange97b} 
\begin{equation}
\frac{dE}{dt} =
\frac{d}{dt}\Big(2K - v\frac{\partial \mathcal{L}}{\partial v}\Big)
- \frac{\partial \mathcal{L}}{\partial t}.
\label{dEdt2}
\end{equation}
Thus two conditions are needed for $E$ to be conserved:
$\partial \mathcal{L}/\partial t = 0$ and 
$v\partial \mathcal{L}/\partial v = 2K$. The second 
condition is equivalent to the requirement $E = \mathcal{H}$.

By using energy conservation, the squared velocity at position $x$ is 
found to be\cite{Hamel49,Calkin89,Marion95,Thornton04}
\begin{equation}
v^2 = 2gx\frac{1-\frac{1}{2}\hat{x}}{1-\hat{x}}, 
\label{v2}
\end{equation}
where $\hat{x} = x/L$. A Taylor expansion for small $\hat{x}$, 
\begin{equation}
\dot{x}^2 \approx 2gx [1 
+ \frac{1}{2}(\hat{x} + \hat{x}^2 + \ldots ) ], 
\label{v2Expanded}
\end{equation}
shows that the falling chain falls faster than free fall right 
from the beginning. Its falling speed then increases monotonically beyond
free fall, and reaches infinity as $\hat{x} \rightarrow 1$. 

We can obtain from Eq.~(\ref{Lagrange}) Lagrange's equation of 
motion for the falling folded chain:
\begin{equation}
m_{R}g - \frac{1}{4}\mu v^2 = \dot{p}_{R} 
= m_{R}\dot{v} + \dot{m}_{R}v.
\label{LEoM}
\end{equation}
We can then verify by direct substitution that the energy-conserving
solution (\ref{v2}) satisfies Eq.~(\ref{LEoM}). Equation~\eqref{LEoM} 
can also be solved directly for $v^2$ by using the identity
\begin{equation}
\dot{v} = \frac{1}{2} \frac{dv^2}{dx}
\label{dotv}
\end{equation}
to change it into a first-order inhomogeneous differential equation for $v^2(x)$.

Lagrange's equation (\ref{LEoM}) is particularly helpful in understanding 
the problem conceptually because it uniquely defines the chain tension $-\mu v^2/4$ that acts upward on the bottom of the right arm at the 
point $B_{R}$ shown in Fig.~\ref{fig:bungee}. This tension comes from the $x$ 
dependence of the kinetic energy and serves the important function of 
enforcing energy conservation. The mistake made in the erroneous 
energy-nonconserving solution is to omit this term. We 
shall explain in the next section why this tension points up and not down, 
as might be expected naively.

It is interesting to apply our analysis to a chain falling from 
a resting heap on a table through a hole in it because this situation 
is even more transparent. Let $x$ be the falling distance, 
now measured from the table. The falling chain is described by
\begin{subequations}
\begin{align}
\mathcal{L}(x,v) &= \frac{\mu}{2}xv^2 + \mu g\frac{x^2}{2} \\
p_x &= \frac{\partial \mathcal{L}}{\partial v} = \mu x v \\
\mathcal{H} &= \frac{p_x^2}{2\mu x} - \frac{\mu g x^2}{2} = E, 
\label{LpH}
\end{align}
\end{subequations}
where the subscript $x$ refers to the falling part of the chain
of length $x$. Because the Lagrangian $\mathcal{L}$ is not explicitly 
time-dependent, we again find $\partial \mathcal{L}/\partial t = 0$ 
and a conservative system. Energy conservation can be 
written in the form 
\begin{equation}
E = \frac{1}{2}\mu x(v^2-gx) = 0.
\label{E=0}
\end{equation}
The resulting solution,\cite{Sousa04} 
\begin{equation}
v^2 = gx,
\label{v2Heap}
\end{equation}
shows that the acceleration of the falling chain is $g/2$, not the value 
$g/3$ of Cayley's energy-nonconserving chain.

The reason for the difference can be seen in Lagrange's equation 
of motion 
\begin{equation}
m_xg + \frac{1}{2}\mu v^2 = \dot{p}_x 
= m_x\dot{v} + \dot{m}_xv.
\label{LEoM2}
\end{equation}
In the incorrect treatment, the downward tension $\mu v^2/2$ that comes from 
the $x$ dependence of the kinetic energy of the falling chain is missing. 

With or without the chain tension term, 
the differential equation~\eqref{LEoM2} describes a
system undergoing a constant acceleration 
$\dot{v} = a$. Hence $v^2 = 2ax$. The differential equation can then be
reduced term by term to the algebraic equation,
\begin{align}
g + sa &= a + 2a, \label{AlgEq.1}\\
\noalign{\noindent giving}
a &= \frac{g}{3-s}
\label{AlgEq.2}
\end{align}
A switching function $s=1$ or 0 has been added to the second term 
on the left in Eq.~\eqref{AlgEq.1}. Hence the
solution is
$a = g/2$ for $s=1$ with the chain tension, and $a= g/3$ for $s=0$ without
the chain tension.

We see that the Lagrangian approach gives a straightforward way of
generating the correct equations of motion in a situation that is
confusing.

\section{Mass transfer between subchains}
\label{sect:massX}

We now clarify how the falling chain transfers mass from 
one subchain to the other. Assume that subchain 2 of mass 
$m_2 + \Delta m$ and velocity ${\bf v}_2$ transfers a small mass $\Delta m$ 
at velocity {\bf u} to subchain 1 of mass $m_1 - \Delta m$ and velocity 
${\bf v}_1$. The transferred mass is related to the subchain masses as
\begin{equation}
\Delta m = \Delta m_1 = - \Delta m_2.
\label{dm}
\end{equation}
At the receiving subchain 1, the initial and final momenta are
\begin{subequations}
\label{m1}
\begin{align}
{\bf p}_{1i} &= (m_1 - \Delta m){\bf v}_1 + {\bf u}\Delta m,
\\
{\bf p}_{1f} &= m_1({\bf v}_1 + \Delta {\bf v}_1),
\end{align}
\end{subequations}
where we have included the momentum of the transferred mass $\Delta m$ 
in the initial state, for the sake of notational simplicity. The total 
momentum change,
\begin{equation}
\Delta {\bf p}_1 = {\bf p}_{1f} - {\bf p}_{1i} 
= m_1 \Delta{\bf v}_1 + \Delta m ({\bf v}_1 - {\bf u}),
\label{dp1}
\end{equation}
on receiving the transferred mass $\Delta m$ can be associated with 
an impulse ${\bf F_1} \Delta t$ received from an external force 
\begin{equation}
{\bf F}_1 \equiv \frac{d{\bf p}_1}{dt} = \frac{d}{dt}(m_1{\bf v}_1) 
- {\bf u}\frac{dm_1}{dt}.
\label{F1}
\end{equation}
This variable mass equation of motion holds whether or not 
the system is conservative. 

In a similar way, we can show that subchain 2 on emitting the 
transferred mass experiences an external force
\begin{equation}
{\bf F}_2 \equiv \frac{d{\bf p}_2}{dt} = \frac{d}{dt}(m_2{\bf v}_2) 
- {\bf u}\frac{dm_2}{dt}.
\label{F2}
\end{equation}
Note how these well-known ``rocket'' equations take the same form 
whether the rocket is discharging or absorbing masses.
Because the total chain mass $M= m_1 + m_2$ is constant, the sum of 
these variable mass equations is just the simple equation
\begin{equation}
{\bf F} = {\bf F}_1 + {\bf F}_2 = \dot{\bf P}, 
\label{F}
\end{equation}
for the center of mass of the entire chain. The internal
forces due to mass transfer always cancel out for any choice of 
{\bf u} when the total mass $M$ is constant.\cite{Lagrange97c}

The velocity {\bf u} of the mass transfer is not arbitrary, however. 
It too is determined uniquely by the chain tension term in Lagrange's 
equation of motion. For the falling folded chain, the second term on 
the left-hand side of Eq.(\ref{LEoM}) gives
\begin{equation}
u\frac{dm_{R}}{dt} = -\frac{1}{4}\mu v^2,
\label{uBungee}
\end{equation}
and for the chain falling from a resting heap, the second term on 
the left-hand side of Eq.(\ref{LEoM2}) gives
\begin{equation}
u\frac{dm_x}{dt} = \frac{1}{2}\mu v^2.
\label{uHeap}
\end{equation}
Thus $u = v/2$ for both falling chains. The two chains differ in that 
the falling folded chain has a fold in it, suggesting that the 
fold falling with the speed $u = v/2$ is the natural location of mass 
transfer. For the chain falling from a heap on a table, on the other hand, 
the mass transfer takes place at a table edge whether sharp or rounded, 
but it is not obvious what the velocity of the transferring link is at the 
moment of the transfer. The answer from Lagrange's equation of 
motion is that it is also the mean velocity ${\bf u} = ({\bf v}_1 + 
{\bf v}_2)/2$ of the two subchains.

We now show that this mean velocity for mass transfer is not accidental,
but is required for energy conservation. To simplify the situation, 
consider a mass transfer that occurs
instantaneously at the same height so that the gravitational force is
not involved. Momentum is then conserved at each subchain. For the
receiving subchain, momentum conservation gives 
$p_{1f} = p_{1i}$, or
\begin{equation}
\Delta {\bf v}_1 = \frac{\Delta m_1}{m_1}({\bf u} - {\bf v}_1).
\label{dv1}
\end{equation}
The receiving process at subchain 1 is a totally inelastic collision that 
involves a kinetic energy change of
\begin{eqnarray}
\Delta K_1 &=& K_{1f} - K_{1i} 
\nonumber \\
&=& \frac{m_1}{2}({\bf v}_1+\Delta{\bf v}_1)^2 - \frac{m_1-\Delta m_1}{2}v_1^2 
- \frac{\Delta m_1}{2} u^2 \nonumber \\
&\approx& -\frac{\Delta m_1}{2} ({\bf u}-{\bf v}_1)^2
\label{dK1}
\end{eqnarray}
to the leading order in $\Delta m_1/m_1$. Here the kinetic energy $K_{1i}$
in the initial state includes the kinetic energy $u^2\Delta m_1/2$ of the
absorbed mass $\Delta m_1$. The net change in kinetic energy is just Carnot's
energy loss.\cite{Gillispie71}

At the emitting subchain 2, momentum is also conserved, thus giving
\begin{equation}
\Delta {\bf v}_2 = \frac{\Delta m_2}{m_2}({{\bf u} - \bf v}_2).
\label{dv2}
\end{equation}
The resulting kinetic energy change can be shown to be
\begin{equation}
\Delta K_2 = K_{2f} - K_{2i} 
\approx -\frac{\Delta m_2}{2} ({\bf u}-{\bf v}_2)^2.
\label{dK2}
\end{equation}
Again the final kinetic energy $K_{2f}$ at subchain 2 includes the kinetic 
energy $u^2\Delta m/2$ of the emitted mass $\Delta m$. 

An examination of these results shows that the total change in the kinetic 
energy $\Delta K = \Delta K_1 + \Delta K_2$ vanishes only when the mass 
$\Delta m = \Delta m_1 = - \Delta m_2$ is transferred at the mean
velocity ${\bf u} = ({\bf v}_1 + {\bf v}_2)/2$. In other words, 
conservation of kinetic energy is enforced when ${\bf u}$ has this 
mean value. The mass emission step is then the exact time reverse of the 
mass absorption step. Hence the kinetic 
energy is conserved for the entire emission-absorption process. 
(The kinetic energy is also conserved for any {\bf u} when 
${\bf v}_1 = {\bf v}_2$, but this solution is of no interest in our problems.) 

Conversely, because we already know that the falling chains are 
conservative systems, we can conclude that the mass transfer must have taken 
place elastically at the mean velocity ${\bf u} = ({\bf v}_1 + {\bf v}_2)/2$ 
even without actually examining Lagrange's equation. Thus knowledge of 
energy conservation alone allows us to conclude that Cayley's assumption of 
inelastic impacts\cite{Cayley57} is incorrect. 

A related result occurs in elastic collisions where the internal
forces are equal and opposite. As a result, ``the kinetic energy lost 
in compression balances exactly the kinetic energy gained in restitution. 
This is sometimes called the third theorem of
Carnot.''\cite{Papastavridis02} Because we know
that the falling chains are conservative systems, it follows that the mass
transfer taken as a whole constitutes a totally elastic collision. 

One final point needs clarification. According to Eq.~(\ref{LEoM}) 
the fold in the chain exerts an upward tension $T_{R} = - \mu v^2/4$ on the 
right arm. The direction of this tension might appear counter-intuitive 
until it is realized that the rocket engine term 
$\dot{m}_Rv = -\mu v^2/2$ on the right-hand side of Eq.~(\ref{LEoM}) 
term can be moved to the left side, the force side, of the equation.
In this position, the term carries a positive sign and represents a 
downward force that dominates the up-pointing tension. When added to 
the force of gravity, these two extra forces together gives a net 
downward force that causes the downward acceleration to exceed $g$.

For the chain falling from a heap, the situation is upside down and 
a time reverse of the falling folded chain. The mass transfer occurs 
at the top where the chain falls down link by link into the moving arm.
The signs of both the chain tension and the rocket engine term are 
opposite to those in the falling folded chain because the falling arm 
is gaining mass. The rocket engine term, $\dot{m}_xv = \mu v^2$,
when moved to the left or force side of Eq.~(\ref{LEoM2}), dominates 
to give a net up-pointing braking force that prevents the falling 
chain from falling as fast as $g$. However, it is the chain tension 
term that pulls the chain down with an acceleration greater than $g/3$. 

\section{The chain tension at the support}
\label{sect:tension}

The chain tension $T$ of the falling folded chain at the support S 
of Fig.~\ref{fig:bungee} can be calculated in the one-dimensional 
continuum model from Eqs.~(\ref{F}) and (\ref{v2}) using $F = Mg - T$.
The result,\cite{Calkin89,Marion95,Thornton04,Meriam02} 
\begin{equation}
T(\hat{x}) = Mg \frac{2 + 2\hat{x} - 3\hat{x}^2}{4(1-\hat{x})}, 
\label{T}
\end{equation}
is a positive monotonic function of $\hat{x}$ that increases to 
$\infty$ as $\hat{x} \rightarrow 1$. 

Calkin and March\cite{Calkin89} studied experimentally the 
tension $T$ of a linked chain with $N=81$ links. They measured a 
maximal tension of 25$Mg$ as the chain approached the bottom. To 
understand this result within a simple theoretical framework, 
we shall assume that the theoretical tension (\ref{T}) of the 
ideal chain with $N \rightarrow \infty$ holds until the last
link remains standing upright. The chain tension at that moment 
is $T(79/81) = 11.1 Mg$. 

The tip of the last link will next fall a distance of 
$2\ell = L-x$, where $\ell = L/N$ is the link length. It does so 
by rotating about a pivot at the contact point between the last 
two links. This rotation can be separated into two steps: first 
a quarter turn to a horizontal position, and then a second quarter
turn to the hanging position at the bottom of its travel. 
To keep the chain center of mass falling straight down, the lower part of the
left arm sways sideways to some maximal displacement after 
the first quarter turn, and then sways back at the second quarter 
turn. This sideway motion will not change the vertical tension. 

In the first quarter turn, the falling chain tip is still above 
the pivot, meaning that fractions of the rotating link are still
coming to rest against the left arm until the last link is
horizontal. Hence the theoretical tension (\ref{T}) can be expected 
to hold until $\hat{x} = 80/81$, where $T$ has almost doubled to
21.2$Mg$. 

In the second and final quarter turn, the chain tip is below the 
pivot. The speed of the chain tip continues to increase, but now 
only by a freely falling rotation. The main consequence of this 
final quarter turn is to convert the vertical velocity $\dot{x}$ 
to a slightly larger horizontal velocity as the chain tip reaches the 
bottom. At that moment, the chain tension $T$ has increased by 
the weight $Mg/N$ of the last link. Because this final increase is 
very small, our simple analysis yields a final result of 
about $21Mg$, in rough agreement with experiment. The final swing 
of the rotating link is easily reproduced by a 
falling chain made up of paper clips.

We believe that the remaining discrepancy comes primarily from 
approximating the linear density $\mu$ of the chain as uniform when 
it is not. The Calkin-March chain appears to be a common or standard 
link chain made up of straight interlinking oval links. At places 
where the links hook into each other, the linear 
density increases by at least a factor of two because all four sides 
of two links appear in cross section instead of the two sides of a 
single link. If we also count the bends of the links, we find a 
significant mass concentration at the linkages. Some of this mass 
concentration at the linkage for the last link should be allowed to 
produce some tension before the last link 
falls down from the horizontal position. Furthermore, this effect 
appears to be larger than any energy loss caused by possible
slippage at the loose linkages of the chain.

The observed maximum chain tension of $25Mg$ can be reproduced at 
$\hat{x} = 0.9896$, an increase of 0.0019 from the theoretical value of
$80/81 = 0.9877$. Each link in the chain has an inside length
of $\ell = 0.97''$. Hence the observed maximum tension is reproduced if
we assume that an additional $0.15''$ of the last link still produces
tension according to the theoretical formula (\ref{T}) after it falls
through the horizontal position.

We note that the link length used in the Calkin-March experiment\cite{Calkin89} 
matches that of the lightest proof coil chain manufactured by the 
Armstrong Alar Chain Corporation,\cite{Armstrong} but that the Armstrong 
chain is too heavy by a factor 1.75. The match would be good if
the material diameter, that is, the diameter of the metal loop in the link, 
is decreased from the Armstrong chain value of $d = (7/32)''$ to
$(5/32)''$. For this estimated matter diameter $d$, the extra length of 
$0.15''$ needed to produce the additional tension is about 1.06$d$. We 
leave it to the reader to determine if this is the correct way to analyze 
the discrepancy and if so, how the result of $1.06d$ can be accounted for 
theoretically.

Our simple interpretation is consistent with the general 
features obtained in the numerical simulation of a falling 
folded chain by Tomaszewski and Pieranski.\cite{Tomas05} They separate 
a chain of length $L=1$\,m into 51 links of uniform linear density joined 
by smooth hinges. They solve the 51 coupled Lagrange's equations 
numerically. They find a maximum velocity of about 21.5\,m/s when the last 
link is falling. In our interpretation, the maximum velocity is expected 
to be $v(50/51) = 22.4$\,m/s, very close to the computed value. 
The numerical solution shows a significant amount of oscillation 
in the stationary left arm when the right arm is falling. This feature is 
not included in the simple one-dimensional treatment using only the falling
distance $x$. The loss of kinetic energy to oscillations in the left arm
has the correct sign to account for the difference between the two
theoretical maximal velocities.

In this connection we note that Calkin and March\cite{Calkin89} did 
not report any dramatic left-arm oscillations in their falling folded 
chain. We also do not find them in a falling folded chain of paper clips. 
A falling ball-chain, on the other hand, does show a wave-like vibration
mostly in the lower half of the rebounding chain. This observed damping 
of the theoretical vibrations expected of the hinged-link model of 
Ref.~\onlinecite{Tomas05} seems to suggest that the loose linkages in the 
physical chains do not transport energy readily to the transverse 
motion of the chain.

\section{The last hurrah}
\label{sect:last} 

For the idealized uniform and inextensible falling folded chain, we find 
its center of mass kinetic energy to be 
\begin{equation}
K_{_{\rm CM}} = MgL \frac{\hat{x}(1-\hat{x})(2-\hat{x})}{8} 
\label{KCM}
\end{equation}
in the one-dimensional continuum model. This CM kinetic energy increases 
from 0 at $\hat{x}=0$ to a maximum value at 
$\hat{x} = 1 - 1/\sqrt{3}$ before decreasing to zero again at 
$\hat{x}=1$. The work done against the chain tension $T=Mg - F$, namely
\begin{equation}
W(X) = \!\int_{L/4}^{X}T(X)dX 
= Mg\Big(X - \frac{L}{4}\Big) - K_{_{\rm CM}}. 
\label{W}
\end{equation}
increases monotonically, reaching $MgL/4$ at $x = L$. 
Given the energy-conserving solution (\ref{v2}) of the one-dimensional 
continuum model where the left arm remains at rest, it is clear that the
change in potential energy given in Eq.(\ref{W}) appears as the 
kinetic energy of the right arm 
\begin{equation}
K_{R} = MgL \frac{\hat{x}(2-\hat{x})}{4}. 
\label{KR}
\end{equation}
Hence the work $W(X)$ done against friction is just
\begin{equation}
W(X) = K_{R} - K_{_{\rm CM}} = K_{\rm int},  
\label{KR}
\end{equation}
the internal kinetic energy of the falling arm not already included in 
$K_{_{\rm CM}}$. In a more detailed model where the motion of the 
left arm is also allowed, the excitation energy of the left arm will
have to be included in the energy balance. The resulting $v^2$ 
will then differ from the value given in Eq.(\ref{v2}) for the 
one-dimensional continuum model.\cite{Tomas05} 

At the moment the falling tip of the ideal one-dimensional chain 
turns over and straightens against the resting left arm, even this
internal kinetic energy vanishes as the entire chain comes 
to rest at full extension. This resting state too has a simple
explanation that is worth repeating: The act of straightening can be
visualized as a completely inelastic Carnot collision in which the
remaining mass 
$\Delta m = m_{R}$ of the right arm is transferred to the left arm 
of mass $m$. Momentum conservation in the laboratory requires that
\begin{equation}
p_f = (m + \Delta m)\Delta v = v\Delta m = p_i.
\label{MomCons}
\end{equation}
The resulting kinetic energy change in this totally inelastic
collision is 
\begin{equation}
\Delta K_{\rm coll} = K_f - K_i 
= - \frac{1}{2}v^2\Delta m \Big( \frac{m}{m + \Delta m} \Big).
\label{KELossLab}
\end{equation}
This analysis shows that in the limit $x \rightarrow L$ when the
right-arm mass vanishes, all its remaining kinetic energy $K_{R} = MgL/4$
is converted into the internal potential energy of the momentarily 
resting chain in a single inelastic collision. For a perfectly 
inextensible chain suspended from a rigid support, $\Delta v$ must 
vanish, which means that the appropriate $m$ must be infinite, 
including not only the finite mass of the left arm but also the 
infinite mass of the support. 

This description is not the end of the story for an actual falling folded
chain. If the chain is an ideal spring, it will be stretched by an
amount consistent with overall energy conservation as the final mass
transfer takes place. This stored potential energy will be used to give the 
chain its kinetic energy on rebound. In actual chains the final rebound 
that follows Carnot's energy loss should also appear, even though the rebound 
is not completely elastic. This grand finale is easily reproduced for a falling 
folded chain of paper clips. 

\section{Conclusion}
\label{sect:conclusion}

We conclude by paying homage to the genius of Lagrange whose 
formulation of classical mechanics helps us to decide definitively if 
a mechanical system is conservative. We have found that Lagrange's 
equation of motion contains a unique description of what 
happens when masses are transferred between the two parts of a falling 
chain, a description that actually enforces energy conservation in 
the falling chain.

Joseph Louis Lagrange (1736--1813) was born Giuseppe Lodovico 
Lagrangia\cite{StAndrews} in Turin of Italian-French parents. He 
introduced purely analytic methods to replace the cumbersome 
geometrical arguments then commonly used in calculus. Using this 
algebraic method, he and his contemporary Leonhard Euler founded 
the calculus of variations as a special branch of mathematics where a 
function that minimizes an integral is to be constructed.\cite{Kline72} 
In his masterpiece {\it M\'ecanique Analytique} (1788),\cite{Lagrange97} 
Lagrange discarded Newton's geometrical approach and recast all of 
mechanics in algebraic form in terms of generalized coordinates whose 
motion satisfies a variational principle, the principle of virtual work. 
He emphasized in the preface that ``No figures 
will be found in this work \ldots only algebraic 
operations \ldots''\cite{Struik67,Lagrange97d} 
He was one of the greatest mathematicians of the 18th century, perhaps 
its greatest.\cite{Ball60a} Truesdell, an admirer of Euler, faults the 
Lagrangian formulation for excessive abstractness that ``conceals the 
main conceptual problems of mechanics.''\cite{Truesdell68} However we have
seen in this paper how Lagrange's method gives definitive answers with
unmatched ease, clarity, and elegance.

\clearpage

\section*{Figure Caption}

\end{document}